\documentclass[pra,amsmath,amsfonts, amssymb,twocolumn,10pt,letterpaper,tightenlines,nofootinbib]{revtex4}

\usepackage{hyperref}

\def\ket#1{\lvert #1 \rangle}
\def\bra#1{{\langle #1 \rvert  }}
\def\tr{{\rm{Tr}}}
\def\det{{\rm{Det}}}

\newcommand{\braket}[2]{\langle #1 \vert #2 \rangle}
\newcommand{\arxiv}[2][quant-ph]{\href{http://arxiv.org/abs/#1/#2}{\tt #1/#2}}

\def\zp{\mathbb{Z}_p}
\def\zd{\mathbb{Z}_d \times \mathbb{Z}_d}
\def\sl{{\sf SL}(2,p)}
\def\slzp{\sl \ltimes \zp^2}
\def\esl{{\sf ESL}(2,p)}
\def\eslzp{\esl \ltimes \zp^2}

\def\sld{{\sf SL}(2,\bar{d})}
\def\slzd{\sld \ltimes \mathbb{Z}_d^2}
\def\esld{{\sf ESL}(2,\bar{d})}
\def\eslzd{\esld \ltimes \mathbb{Z}_d^2}

\def\R#1{#1\textup{\,R\,}p}
\def\N#1{#1\textup{\,N\,}p}

\newcommand{\Lp}[1]{\left(\frac{\displaystyle{#1}}{\displaystyle{p}}\right)}

\renewcommand{\mod}{\bmod}

\newtheorem{theorem}{Theorem}

\newtheorem{definition}{Definition}
\newtheorem{conjecture}{Conjecture}
\newtheorem{corollary}{Corollary}
\def\Pr{{\bf Proof: }}

\begin{document}

\title{On SIC-POVMs in Prime Dimensions}
\author{Steven T. Flammia}
\email{sflammia@unm.edu}
\affiliation{Department of Physics and Astronomy, University of New Mexico, Albuquerque, New Mexico 87131}
\date{May 4, 2006}

\begin{abstract}
The generalized Pauli group and its normalizer, the Clifford group, have a
rich mathematical structure which is relevant to the problem of
constructing symmetric informationally complete POVMs (SIC-POVMs).  To
date, almost every known SIC-POVM fiducial vector is an eigenstate of a ``canonical''
unitary in the Clifford group.  I show that every canonical unitary in prime
dimensions $p > 3$ lies in the same conjugacy class of the Clifford group
and give a class representative for all such dimensions.  It follows
that if even one such SIC-POVM fiducial vector is an eigenvector of such a
unitary, then all of them are (for a given such dimension).  I also
conjecture that in \emph{all} dimensions $d$, the number of conjugacy
classes is bounded above by 3 and depends only on $d\mod 9$, and I support this claim with computer computations in all dimensions $< 48$.
\end{abstract}

\maketitle

\section{Introduction}

In the field of quantum information, many diverse applications make
frequent use of the notion of {\it optimal measurement\/}: optimal quantum
state tomography~\cite{QSE}, quantum cloning~\cite{Gisin,Scott},
error-free state discrimination~\cite{Chefles1998a,Chefles1998b}, certain
quantum key distribution protocols~\cite{Renes2004b, Renes2004c}, and
quantum algorithms~\cite{BCvDa, BCvD} are but a few examples.  Often, the
optimal solution to a problem is given by a generalized measurement known
as a {\it positive-operator valued meausure\/}, or POVM~\cite{N&C}.  A
POVM is a set of positive operators $E_i$ such that the probability of
obtaining the $i$th outcome is given by $\tr(E_i \rho)$, where $\rho$ is
the density operator for the system being measured.  A POVM must satisfy
the completeness condition, $\sum_i E_i = 1$, which is equivalent to
saying the probabilities of the outcomes must sum to unity.  In this
paper, we deal only with POVMs having a finite number of elements.

If the statistics of a POVM are sufficient to uniquely determine any
quantum state with fixed dimension $d$, the POVM is said to be {\it
informationally complete\/} (for that particular $d$).  The notion of
informational completeness was first discussed in
Ref.~\cite{Prugovecki1977a}, and subsequently in
Refs.~\cite{Busch1989a,Busch1989b, Hellwig1993,Peres1993a}, as well as in
Refs.~\cite{Flammia05, Finkelstein04} when applied to just pure states.
Informationally complete POVMs have applications to foundational studies
where they play a role in the Bayesian formulation of quantum mechanics
\cite{CFS1, CFS2, FS, Fuchs}, and make particularly nice ``standard
quantum measurements'' \cite{Hardy}.  Since there are $d^2-1$ parameters
in an unknown density operator, an informationally complete POVM requires
at least $d^2-1$ independent measurement outcomes; together with the
completeness condition this implies that a {\it minimal\/} informationally
complete POVM is one with exactly $d^2$ elements~\cite{Weigert05}.  If an
informationally complete POVM is to be maximally efficient at determining
a state via tomography, then the POVM elements should be proportional to
one-dimensional projectors.  If this is the case, and in addition the
vectors onto which the POVM elements project are evenly spaced in Hilbert
space, i.e.~the squared inner products are the same for any pair of
distinct vectors, then the POVM is said to be {\it symmetric\/}.  This
motivates the definition of a symmetric informationally complete POVM, or
SIC-POVM.

\begin{definition}
    A SIC-POVM $\mathcal{S}$ on a $d$ dimensional Hilbert space
$\mathbb{C}^d$ is a POVM with $d^2$ elements $E_i$ such that each $E_i \in
\mathcal{S}$ is rank one, i.e.~$E_i \propto \ket{\psi_i}\!\bra{\psi_i}$
for some $\ket{\psi_i} \in \mathbb{C}^d$, and each pair of distinct normalized vectors satisfies
\vspace{-.5em}\begin{equation} \label{E:sicdef}
    \left|\braket{\psi_i}{\psi_j}\right|^2 = \frac{1}{d+1}\ .
\end{equation}
\end{definition}

Thus, a SIC-POVM is a POVM that is informationally complete, minimal, and
symmetric.  (This is actually redundant because minimal and symmetric
implies informationally complete.)  SIC-POVMs were discovered by Zauner
\cite{Zauner} and independently by Renes et.~al.~\cite{Renes2004a}.  Exact
solutions to Eq. \ref{E:sicdef} exist in dimensions 2-13,15 and 19, and
numerical examples exist in all dimensions~$\le 45$
\cite{Zauner,Renes2004a,Grassl,Grassl2,GrasslP,Appleby,Hoggar}.  
SIC-POVMs are known in
the mathematical literature as equiangular lines, and have been studied
for a number of years in the context of frame theory, $t$-designs, and
spherical codes \cite{Delsarte}.

A POVM is {\it group covariant\/} \cite{DAriano} if there exists a group
$G$ of order $d^2$ with a projective unitary irreducible representation
(UIR) on $\mathbb{C}^d$ such that the conjugation action of the projective
UIR on the POVM merely permutes the measurement outcome labels.  Nearly
every SIC-POVM to date has been constructed using group covariance under
the group $\zd$ in a manner defined as follows \cite{OtherGroups}.  
Fix an orthonormal basis for $\mathbb{C}^d$, and define the
operators
\begin{equation}
    D_{jk} = \omega^{jk} \sum_{n=0}^{d-1} \omega^{j n} \ket{n \oplus k}\bra{n} ,
\end{equation}
where $\omega = e^{2 \pi i/d}$ is a primitive $d$th root of unity and
$\oplus$ denotes addition mod $d$.  The operators $D_{jk}$ form a
projective UIR of $\zd$ and generate the {\it generalized Pauli group\/},
or GP group, denoted $GP(d)$.  Then construct a SIC-POVM by finding a
normalized {\it fiducial vector\/}, $\ket{\psi_0}$, such that the set of
distinct vectors in $\left\{D_{jk}\ket{\psi_0}\right\}_{j,k=0}^{d-1}$ have the same absolute inner product onto the fiducial state.  This implies
Eq.~\ref{E:sicdef}, and the SIC-POVM is then formed by
the set of subnormalized projectors
\begin{equation}
    E_{jk} = \frac{1}{d}D_{jk}\ket{\psi_0}\bra{\psi_0} D^\dag_{jk}.
\end{equation}
In this paper, we are interested solely in SIC-POVMs formed via this
construction; for the rest of the paper, ``SIC-POVM'' and ``fiducial
vector'' imply GP covariance.

Since the SIC-POVMs we consider are all covariant under the action of
$GP(d)$, we can also consider the action of the normalizer of $GP(d)$ in
$\sf{U}(d)$, the so-called {\it Clifford group\/}, denoted $C(d)$.  Given
any fiducial vector $\ket{\psi_0}$ and a Clifford group element $U$,
$U\ket{\psi_0}$ is also a fiducial vector.  We can extend $C(d)$ to allow
anti-unitary operators as well, obtaining the {\it extended Clifford
group\/}, denoted $EC(d)$.  Then given a fixed fiducial vector
$\ket{\psi_0}$, every SIC-POVM in that orbit can be written as $U\ket{\psi_0}$ for some
$U \in EC(d)$.  Since the action of $C(d)$ or $EC(d)$ on the SIC-POVM is a
conjugation action, we are really interested in $C(d)/I(d)$ and
$EC(d)/I(d)$, where $I(d)$ is the center of $U(d)$ consisting of all
matrices which are just a phase times the identity matrix.  We denote
these projected groups as $PC(d)$ and $PEC(d)$, respectively.

We now mention a theorem due to Appleby \cite{Appleby} which characterizes
the groups $PC(d)$ and $PEC(d)$.  Since we are primarily concerned with
prime dimensions $>3$ in this paper, we will state the theorem restricted
to this special case.  Recall that the group $\sl$ is the group of $2
\times 2$ matrices defined over the field $\zp$ having unit determinant in
$\zp$.  Define $\esl$ to be the group obtained by adding the generator $J=
\begin{pmatrix}
1&0\\
0&-1
\end{pmatrix}
$ to $\sl$.
\begin{theorem}[Appleby]\label{T:appleby}
Let $p$ be a prime $> 3$.  Then $PC(p)$ is isomorphic to $\slzp$, and
$PEC(p)$ is isomorphic to $\eslzp$.
\end{theorem}

Before we can appreciate the significance of this theorem for our
purposes, we need one more definition.  Define the {\it Clifford trace\/}
of any element $U \in PEC(p)$ as follows.  From Theorem~\ref{T:appleby},
there exists an isomorphic image of $U$ in $\eslzp$ which we can represent
as an ordered pair $(F,\chi)$, where $F \in \esl$ and $\chi \in \zp^2$.
The Clifford trace of $U$, denoted $\tr_C(U)$ is defined as
$\tr_C(U)=\tr(F)$, where the trace on the right-hand side is taken over
$\zp$.  Following Appleby \cite{Appleby}, we call any $U$ with $\tr_C(U) = -1 \mod p$ a {\it canonical element\/}, provided it is not the identity (which can only happen when $p=3$).  As an example of such an element that exists in every finite dimension, define the matrix $Z = \begin{pmatrix}
	0&-1\\
	1&-1
\end{pmatrix}$.  This matrix, whose importance was first recognized by Zauner \cite{Zauner}, will feature prominently in the main result of this paper.  (Ref.~\cite{CCDG} also discusses an element of $\sld$ that is conjugacy equivalent to $Z$ and mentions its importance to the SIC-POVM problem.)

The following three conjectures relate Theorem \ref{T:appleby} to SIC-POVMs
through the Clifford trace \cite{Appleby, Zauner}.  All three conjectures assert that a SIC-POVM exists in every finite dimension, but they differ in the properties of the fiducial vectors used to generate the SIC-POVM.  Since we are primarily interested in the case of prime dimensions ($p>3$), we state the conjectures specialized to this case and refer the reader to Ref.~\cite{Appleby} for a discussion of the more general conjectures.

\begin{conjecture}[Appleby]  SIC-POVMs exist for every prime dimension, and every SIC-POVM fiducial vector is an eigenvector of a canonical element of $PC(p)$.
\end{conjecture}

\begin{conjecture}[Zauner]
For every prime dimension, there exists a SIC-POVM fiducial vector that is an eigenvector of the unitary operator associated with the matrix $Z$.
\end{conjecture}

\begin{conjecture}[Appleby]
SIC-POVMs exist for every prime dimension, and every SIC-POVM fiducial vector is an eigenvector of a canonical element of $PC(p)$ that is conjugate to the matrix $Z$.
\end{conjecture}

Conjectures 1 and 2 hold for every known 
SIC-POVM, and in fact a further extension to all dimensions (not just primes)
also holds \cite{Appleby}.  Conjecture 3 is clearly stronger than Conjecture 2, and it also implies Conjecture 1.  Grassl \cite{Grassl2} has constructed a counterexample in dimension 12 to the analog of Conjecture 3 extended to composite dimensions, but there are no known counterexamples in other dimensions.  Although Conjecture 3 is not true in general, it is important to know for which dimensions it is valid, as the following illustrates.

Because $EC(d)$ acts on $GP(d)$ via conjugation, if one were to search
for a SIC-POVM by assuming Conjecture 1, it is sufficient
to choose one element from each of the conjugacy classes of $EC(d)$
having Clifford trace $=-1$, and search the (degenerate) eigenspaces of
these elements.  This procedure would yield either a SIC-POVM or (if the search was exhaustive) a counterexample to the conjecture.  

The main result of this paper is to show that such a search as described
above need only check \emph{one} conjugacy class element if the dimension is a prime $>3$, thus reducing the search to one over a bounded number of conjugacy classes.  This is done by demonstrating the equivalence of all three Conjectures when the dimension is prime.
This also shows that if one fiducial vector can be found as an eigenvector
of the canonical class representative ($Z$), then every other fiducial vector in
prime dimensions $>3$ automatically satisfies Conjecture 1.

Before stating the main result in section \ref{S:mainproof}, we discuss
some background results from number theory and prove some theorems
applicable to the proof of the main theorem.  Readers well-versed in
number theory may skip section \ref{S:numbertheory} and proceed directly
to section \ref{S:mainproof}, although it may be useful to skim the former
to glean the notation used in the latter.  In section \ref{S:conj}, we
state an extension of the main theorem and offer supporting numerical
evidence.

\section{Background Results from Number Theory}\label{S:numbertheory}

In this section we introduce a basic concept from number theory, the
Legendre symbol, and state some properties and theorems that will be used
in the proof of the main theorem.  The basic material can be found in any
textbook on the subject (see for example Refs.~\cite{KR97, Andrews71}),
but we review it here for completeness.

Let $p$ be an odd prime and $n$ be any integer such that $\gcd(n,p)=1$.
Then $n$ is a {\it quadratic residue} mod $p$ if there exists an integer
$k$ such that $k^2 = n \mod p$.  If no such integer exists, then $n$ is
said to be a {\it quadratic nonresidue}.  Since we will only be dealing
with quadratic residues mod $p$, we will frequently omit the $p$ and the
word quadratic and simply say, for example,  ``$n$ is a residue'', with
$p$  and quadratic being understood from the context.  We use the symbols
$\R n$ and $\N n$ to denote that $n$ is a residue or nonresidue
respectively.

The {\it Legendre symbol}, $\displaystyle{\Lp{n}}$, is defined by
\begin{equation}
    \Lp{n} =
    \left\{\begin{array}{cll}
        +1 & \mbox{if } \R n, \\
        -1 & \mbox{if } \N n,\\
        0 & \mbox{if } p|n \,.
    \end{array}\right.
\end{equation}

\begin{theorem}\label{T:prop}
Let $m$ and $n$ be any integers, and $p$ an odd prime.  Then the following
properties of the Legendre symbol hold:
$$
\begin{array}{lll}
    \textup{Property 1:} & \vspace{.5em} & \displaystyle{\Lp{n}=n^{(p-1)/2} \!\!\mod p},\\
    \textup{Property 2:} & \vspace{.5em} & \displaystyle{\Lp{mn} = \Lp m \Lp n} ,\\
    \textup{Property 3:} & \vspace{.5em} & \displaystyle{\Lp{n^{-1}}=\Lp{n}},\\
    \textup{Property 4:} & \vspace{.5em} & \displaystyle{\sum_{n=1}^{p-1} \Lp n = 0} ,\\
    \textup{Property 5:} & \vspace{.5em} & \displaystyle{\Lp{-3}}  =
    \left\{\begin{array}{ll}
        +1 & \mbox{\textup{if} } p = +1 \mod 3 ,\\
        -1 & \mbox{\textup{if} } p = -1 \mod 3 .
    \end{array}\right.
\end{array}
$$
\end{theorem}
\Pr As these properties are very basic, their proofs are not particularly
enlightening, so we omit them.  See Refs.~\cite{KR97, Andrews71} for
proofs.  Property 1 is known as {\it Euler's criterion}.  Property~4
simply says that the number of residues and non-residues is exactly
$(p-1)/2$.  $\Box$

We now prove some useful results that we will need in section
\ref{S:mainproof}.  In the interest of brevity the proofs are concise, but
expanded versions of Theorems \ref{T:nn+1} and \ref{T:Np} can be found in
Ref.~\cite{Andrews71}.

\begin{theorem}\label{T:nn+1}
\begin{equation}
    \sum_{n=1}^{p-2} \Lp{n} \Lp{n+1} = -1 .
\end{equation}
\end{theorem}
\Pr Since all integers in the interval $[1,p-2]$ are invertible, we can
``factor'' an $n$ out of the second factor in the sum, using Property 2 to
combine this $n$ with the first factor.
\begin{eqnarray}
    \sum_{n=1}^{p-2} \Lp{n} \Lp{n+1} &=& \sum_{n=1}^{p-2} \Lp{n^2} \Lp{1+n^{-1}} \nonumber \\
    &=& \sum_{n=1}^{p-2} \Lp{1+n^{-1}}.
\end{eqnarray}
Because all the inverses of elements in the range $[1,p-2]$ are still in
that range, this sum has the same value as the following sum, which can be
immediately evaluated by reindexing the sum and using Property 4.
\begin{equation}
    \sum_{n=1}^{p-2} \Lp{1+n^{-1}} = \sum_{n=1}^{p-2} \Lp{1+n} = -1.
\end{equation}
$\Box$

\begin{theorem}\label{T:Np}
Let $N(p)$ be the number of consecutive residues in the interval
$[1,p-1]$.  Then $N(p)$ is given exactly by
\begin{equation}
    N(p) = \frac{1}{4} \left(p-4-(-1)^{(p-1)/2}\right) .
\end{equation}
\end{theorem}
\Pr The proof follows Ref.~\cite{Andrews71}.  Let the function $c_p(n)$ be
defined by
\begin{equation}
c_p(n) = \left\{
\begin{array}{ll}
    1 & \mbox{if } \R n \mbox{ and } \R{(n+1)}, \\
    0 & \mbox{otherwise.}
\end{array}
\right.
\end{equation}
Thus $c_p(n)$ is the indicator function for adjacent residues.  Note that
\begin{equation}
    c_p(n) = \frac{1}{4}\left(1+\Lp n\right)\left(1+\Lp{n+1}\right) .
\end{equation}
Then we can write $N(p)$ as
\begin{equation}
    N(p) = \sum_{n=1}^{p-2} c_p(n) .
\end{equation}
Expanding the expression for $c_p(n)$, we get four sums:
\begin{equation}
    N(p) = \frac{1}{4} \sum_{n=1}^{p-2} \left(1+\Lp n+\Lp{n+1}+\Lp n\Lp{n+1} \right).
\end{equation}
The first three can be evaluated using Euler's criterion and Property 4,
while the last is the content of Theorem \ref{T:nn+1}.  The result follows
directly. $\Box$

\begin{theorem}\label{T:ressum}
\begin{equation}
    \sum_{\R n}\Lp{n+1} = \frac{(1-p)}{2}+2 N(p) + \frac{1+(-1)^{(p-1)/2}}{2}=-1 .
\end{equation}
\end{theorem}
\Pr Since there are exactly $(p-1)/2$ residues, the {\it least\/} possible
value of this sum is achieved if every term is $-1$, giving the first term
in the middle equality.  However, this lower bound under counts whenever
both $n$ and $n+1$ are residues, so we add $2 N(p)$ to correct for this.
The only other consideration is if $\R{-1}$, a term which is not included
in the $N(p)$ correction, since $0$ is neither a residue nor a nonresidue.
In this case, we should add only 1 instead of two, since the Legendre
symbol of 0 is 0.  The final term
\begin{equation}
    \frac{1}{2} (1+(-1)^{(p-1)/2})
\end{equation}
has the requisite property.  Summing these terms and plugging in the
formula from Theorem~\ref{T:Np} completes the proof. $\Box$

\begin{theorem}\label{T:upsilon}
Let $f(x)$ be a polynomial with integral coefficients.  Let $\Upsilon(f)$
be the number of mutually incongruent solutions in $x$ and $y$ to the
equation $y^2 = f(x) \mod p$.  Then
\begin{equation}
    \Upsilon(f) = p+\sum_{n=0}^{p-1} \Lp{f(n)} .
\end{equation}
\end{theorem}
\Pr If $\R{f(n)}$, then there are two solutions, $\pm y$.  If $\N{f(n)}$,
there are no solutions, and if $f(n)=0$, there is only one solution,
$y=0$.  We simply note that the following term counts the number of
solutions correctly for fixed $n$, and the proof is immediate.
\begin{equation}
    \left(1+\Lp{f(n)}\right) =
    \left\{
    \begin{array}{ll}
        2 & \mbox{if } \R{f(n)}  \\
        0 & \mbox{if } \N{f(n)} \\
        1 & \mbox{if } f(n) = 0 .
    \end{array}
    \right.
\end{equation}
$\Box$

\section{All Canonical Unitaries are Conjugacy Equivalent}\label{S:mainproof}

In this section we prove the main theorem.  Throughout this section,
assume that $p$ is a prime $> 3$.  Because of the isomorphism in Theorem
\ref{T:appleby}, we can work exclusively in $\eslzp$.  In fact, we need
only work in $\slzp$ because SIC-POVMs always come in complex conjugate
pairs; any fiducial vector which is an eigenvector of an element in
$PEC(d)$ that is not an eigenvector of an element of $PC(d)$ will have a
conjugate fiducial vector which \emph{is} an eigenvector of an element of
$PC(d)$.  So a search for a fiducial vector satisfying 
Conjecture 1 need only check elements of $\slzp$.  Recall that the
composition rule on $\slzp$ is defined as follows:
\begin{equation}
    (F,\chi) \circ (G,\zeta) = (FG,\chi+F \zeta) .
\end{equation}
The first step is to prove that one need only consider elements of the
form $(F,0)$, which we prove as a separate theorem.

\begin{theorem}\label{T:nochi} For all $(F,\chi ) \in \slzp$ with
$\tr(F)\not=2 \mod p$, $(F,\chi)$ is in the same conjugacy class as
$(F,0)$.
\end{theorem}
\Pr We would like to show that there always exists $(G,\zeta) \in \slzp$
such that
\begin{equation}
    (G,\zeta)\circ (F,\chi) \circ (G,\zeta)^{-1} = (F,0) .
\end{equation}
To satisfy this conjugacy relation, we will see that it is sufficient to consider elements with $G=I$.
Expanding the previous formula with $G=I$, we obtain an equation relating
$\zeta$ to $F$ and $\chi$.
\begin{equation}
    \chi = (F-I) \zeta .
\end{equation}
This equation can be solved for $\zeta$ whenever $\det(F-I)\not=0 \mod p$.
Expanding the determinant of $F-I$,
we obtain
\begin{equation}
    \det(F)-\tr(F)+1\not=0,
\end{equation}
from which the trace condition on $F$ follows immediately.
$\Box$

The main theorem is concerned with $F$ matrices having trace $=-1\mod p$.
Since the identity matrix satisfies this condition when $p=3$,
i.e.~$\tr(I)=2=-1\mod3$, it is necessary to exclude this case.

Note that in the previous proof, we considered only elements of $\slzp$ of
the form $(I,\zeta)$.  In the next proof, we work only with $G \in \sl$.
By concatenating these two results, our general element is of the form
$(G,\zeta)$.

We now embark on a proof of the main theorem, making use of the results of
Section~\ref{S:numbertheory}.

\begin{theorem}\label{T:main}
Let $p$ be a prime $> 3$, and $F \in \sl$ with $\tr(F) = -1 \mod p$.  Then
there exists a $G \in \sl$ such that
\begin{equation}
    G F G^{-1} = Z = \begin{pmatrix} 0&-1\\ 1&-1 \end{pmatrix} \, .
\end{equation}
\end{theorem}
\Pr Let
\begin{equation}
    F=
    \begin{pmatrix}
        \alpha & \beta \\
        \gamma & -1-\alpha
    \end{pmatrix}\;,\qquad
    G=
    \begin{pmatrix}
        a & b \\
        c & d
    \end{pmatrix}
\end{equation}
 be matrices in $\sl$.
Note that the conditions $\det(F)=-\tr(F)=1$ hold, and we have the freedom
to choose the matrix elements of $G$ as long as they satisfy the
constraint $\det(G)=1$. If the matrix elements $a$ and $b$ of $G$ are
chosen to be
\begin{equation}
    a=c(\alpha+1)+d\gamma \ , \ b=c\beta -d\alpha \, ,
\end{equation}
then the relation
\begin{equation}
    GF= Z G
\end{equation}
always holds, so $c$ and $d$ are free parameters that must be chosen to
satisfy $\det(G)=1$.  Expanding the formula for $\det(G)$ and simplifying,
we obtain the following equation for $c$ and $d$ as a function of the
matrix elements of $F$:
\begin{equation}\label{E:detg1}
    d^2 \gamma + c d (2 \alpha +1)-c^2 \beta =0.
\end{equation}
We must show that this equation always has a solution, a task which takes
up the remainder of the proof.  We proceed in three cases: $\gamma = 0$,
$\R \gamma$, and $\N \gamma$.\\

{\it Case 1:} $\gamma = 0$.  \\

In this case, setting $c=1$, Eq.~\ref{E:detg1} simplifies to
\begin{equation}
    d (2 \alpha +1)=\beta.
\end{equation}
This equation can always be solved for $d$ unless $\alpha = -2^{-1}$.  But
suppose by contradiction that it was possible that $\alpha =-2^{-1}$.
Then comparing with the constraint on the determinant of $F$, we find that
\begin{equation}
    \det(F)=1 \mod p \ \Rightarrow \ -2^{-1}(-1+2^{-1})=1 \mod p\, ,
\end{equation}
which implies that $4=1 \mod p$, something which impossible since
$p\not=3$.  This completes the demonstration of Case 1.

Before proceeding to the second two cases, it pays to simplify the form of
Eq.~\ref{E:detg1} using the assumption that $\gamma \not= 0$.  Using the
fact that $\gcd(2 \gamma, p) =1$, we can complete the square in
Eq.~\ref{E:detg1} while preserving its solutions to obtain
\begin{equation}
    (2\gamma d + c (2\alpha+1))^2=(c(2\alpha+1))^2+4\gamma(1+c^2\beta) .
\end{equation}
Since $\R 4$, so is $\R{4^{-1}}$, and by expanding the right hand side we
can further simplify this to
\begin{equation}
    (4^{-1/2}2\gamma d + c 4^{-1/2} (2\alpha+1))^2=\gamma - 3 (4^{-1}) c^2 .
\end{equation}
Now a simple change of variables given by
\begin{equation}
    x=d \gamma + c (\alpha+2^{-1})\ , \ y= 2^{-1} c
\end{equation}
allows this to be written in the very compact form
\begin{equation}\label{E:xy}
    x^2=\gamma-3y^2 .
\end{equation}
From this simplified form, we can immediately solve Case 2.\\

{\it Case 2:} $\R \gamma$.\\

If $\R \gamma$, simply choose $y=0$ (implying $c=0$) and then
$x=\gamma^{1/2}$ can be inverted for $d$.  This concludes Case 2.

The remaining case is more difficult; it is the reason we developed
so much machinery in section \ref{S:numbertheory}.\\

{\it Case 3:} $\N \gamma$.\\

By Theorem \ref{T:upsilon}, the number of solutions $\Upsilon$ to
Eq.~\ref{E:xy} is given by
\begin{equation}
    \Upsilon = p+ \sum_{n=0}^{p-1} \Lp{\gamma-3n^2} \,.
\end{equation}
By taking out the $n=0$ term from the sum and ``factoring out'' a $\gamma$
from the Legendre symbol, this becomes
\begin{equation}
    \Upsilon = p -1-\sum_{n=1}^{p-1} \Lp{1-3\gamma^{-1} n^2} \,.
\end{equation}
The sum can now be rewritten to go over only the residues, since $n$
appears only to the second power inside the summand.  A factor of two is
necessary to account for both the square roots of the residue.
\begin{equation} \label{E:upres}
    \Upsilon = p -1-2 \sum_{\R n} \Lp{1-3\gamma^{-1} n} \,.
\end{equation}
This is nearly in a form where Theorem \ref{T:ressum} is applicable.  To
get it in such a form, we consider two cases, $p=\pm1 \mod 3$, and
denote the number of solutions in each case as $\Upsilon_\pm$.  First,
note that since $\N \gamma$, the sum in Eq.~\ref{E:upres} can be reordered
and written
\begin{equation}\label{E:nogamma}
    \Upsilon_\pm = p -1-2 \sum_{\N n} \Lp{1-3n} \,.
\end{equation}
From Property 5 in Theorem \ref{T:prop}, we know when $\R{-3}$ or
$\N{-3}$, so Eq.~\ref{E:nogamma} can be reordered to become
\begin{equation}
    \Upsilon_+ = p-1-2 \sum_{\N n} \Lp{n+1} \,,
\end{equation}
\begin{equation}
    \Upsilon_- = p-1-2 \sum_{\R n} \Lp{n+1} \,.
\end{equation}
To calculate $\Upsilon_+$, note the following simple identity:
\begin{eqnarray}
    \sum_{\N n} \Lp{n+1} &=& \sum_{n=1}^{p-1} \Lp{n+1} -\sum_{\R n} \Lp{n+1}
    \nonumber \\
    &=& -1-\sum_{\R n} \Lp{n+1} ,
\end{eqnarray}
where Property 4 of Theorem \ref{T:prop} was used. So the formula for
$\Upsilon_\pm$ becomes
\begin{equation}
    \Upsilon_\pm = p\pm1\pm2 \sum_{\R n} \Lp{n+1} \,,
\end{equation}
Now plug in the results of Theorem \ref{T:ressum} to obtain
\begin{equation}
    \Upsilon_\pm = p\mp1 \,,
\end{equation}
and so the number of solutions is strictly greater than zero. $\Box$

The proof of Theorem \ref{T:main} demonstrates that there is exactly one conjugacy class with trace $=-1\mod p$ in the group $\slzp$ if the dimension $p$ is a prime $>3$.  The consequences for the Conjectures 1,2 and 3 are summarized in the following corollary.  

\begin{corollary}
For prime dimensions $p>3$, Conjectures 1,2 and 3 are equivalent.
\end{corollary}
$\Box$

\section{A Further Conjecture} \label{S:conj}

To state the conjecture, we make use of the extended theorem classifying
the Clifford group in non-prime dimensions found in Ref.~\cite{Appleby}.
Let
\begin{equation}
    \bar{d} =
    \left\{\begin{array}{cll}
        d & \mbox{if } d \mbox{ is odd}, \\
        2d & \mbox{if } d \mbox{ is even} \,.
    \end{array}\right.
\end{equation}
Then the projective Clifford group $PC(d)$ and the projective extended
Clifford group $PEC(d)$ are homomorphic to $\slzd$ and $\eslzd$,
respectively.  The kernel of the homomorphism is an order 8 subgroup
isomorphic to $\mathbb{Z}_2^3$.  See Ref.~\cite{Appleby} for details.

\begin{conjecture} \label{C:conj}
    Let $T_d$ denote the number of conjugacy classes of the group $\sld$
(for $d>1$) having trace $= -1 \mod d$.  Then $T_d$ is exactly given by
\end{conjecture}
\begin{equation}
    T_d =
    \left\{\begin{array}{cll}
        3 & \mbox{if } 3|d \mbox{ and } 9\!\!\not|d, \\
        2 & \mbox{if } 9|d, \\
        1 & \mbox{otherwise} \,.
    \end{array}\right.
\end{equation}

Note the strange interplay between $d$ and $\bar{d}$.  The results of
section \ref{S:mainproof} establish the truth of this conjecture when $d$
is a prime $>3$.  However, the remaining cases are not approachable via a
direct application of the methods found here because of the presence of
zero divisors in arithmetic modulo $d$.  We therefore leave an analytic
demonstration of Conjecture \ref{C:conj} to future work, and instead
establish its plausibility algorithmically.  Using the computer program
GAP, we have established the truth of Conjecture \ref{C:conj} in all
dimensions $<48$.

There are two points now worth emphasizing.  Conjecture \ref{C:conj} attempts to classify exactly for which dimensions the equivalence of the three Conjectures 1,2 and 3 holds.  The answer appears to be ``any dimension not divisible by 3''.  Note that this agrees with the results in Ref.~\cite{Grassl2}.  Second, the computer program GAP does \emph{not} use floating-point arithmetic.  This means that the algorithmic verification of Conjecture \ref{C:conj} in dimensions $<48$ is \emph{exact}.

\section{Conclusion}

We have established that all canonical unitaries in the projective Clifford
group in a prime dimension $>3$ lie in the same conjugacy class.  Thus, if
even one SIC-POVM fiducial vector is an eigenvector of such a unitary,
then all of them are (for a given such dimension).  We have also advanced
a conjecture which would extend this result to all dimensions and offered
computer calculations as evidence supporting it in all dimensions $<48$.  
These results begin to classify for which dimensions the Conjectures 1,2 and 3 are equivalent.

\acknowledgments

Thanks to Andrew Landahl and Seth Merkel for helpful discussions, and to Carl Caves for comments on this manuscript.  Markus Grassl also provided valuable suggestions for improving this paper.  This research was supported by Office of Naval Research Contract No.~N00014-03-1-0426.

Note Added In Proof:  After this paper was accepted for publication, a proof of conjecture \ref{C:conj} was discovered to have been proven in references \cite{Nobbs1} and \cite{Nobbs2}.

\end{document}